\documentclass[12pt]{article}
\usepackage{epsfig}

\def\be{\begin{equation}}
\def\ee{\end{equation}}
\def\ben{\begin{displaymath}}
\def\een{\end{displaymath}}
\def\ba{\begin{array}{c}}
\def\bal{\begin{array}{l}}
\def\ea{\end{array}}
\def\p{\partial}
\begin{document}

\titlepage

 \begin{center}
{\tiny .}

\vspace{.35cm}

{\Large \bf

Horizons of stability in matrix Hamiltonians

%
%
%
%

   }\end{center}

\vspace{10mm}

 \begin{center}

 {\bf Miloslav Znojil}

 \vspace{3mm}
Nuclear Physics Institute ASCR,

 250 68 \v{R}e\v{z}, Czech Republic

{e-mail: znojil@ujf.cas.cz}

\vspace{3mm}

\vspace{5mm}
%

\end{center}


\vspace{5mm}

\section*{Abstract}

Non-Hermitian Hamiltonians $H\neq H^\dagger$ possess the real
(i.e., observable) spectra inside certain specific, ``physical"
domains of parameters $ {\cal D}= {\cal D}(H)$. In general, the
determination of their ``observability-horizon" boundaries $\p
{\cal D}$ is difficult. We list the pseudo-Hermitian real $N$ by
$N$ matrix Hamiltonians for which the ``prototype" horizons $\p
{\cal D}$ are defined by closed analytic formulae.


\newpage

\section{Introduction}

In  Landau's textbook~\cite{Landau} on Quantum Mechanics the
emergence of an instability in a system is illustrated via a
particle in the potential $V(\vec{x})=G/|\vec{x}|^2$. The critical
value $G_{(min)}=-1/4$ of its strength represents a ``horizon"
beyond which the particle starts falling on the center. {\it Vice
versa}, the system remains stable and physical on all the interval
${\cal D}= (-1/4, \infty)$ of couplings $G$. From the pragmatic
point of view the Landau's example is not too well selected since
the falling particle should release, hypothetically, an infinite
amount of energy during its fall. A better example of the loss of
stability is provided by the Dirac's electron which moves in a
superstrong Coulomb potential: In the language of physics,
particle-antiparticle pairs are created in the system beyond a
critical charge ($Z_{(max)}=137$ in suitable
units~\cite{Greiner}).

What is shared by the above two sample Hamiltonians $H(\lambda)$
is that they are well defined in a certain domain ${\cal D}$ of
parameters $\lambda$ while they lose sense and applicability for
parameter(s) beyond certain horizon(s) $\lambda_{(max/min)}$. On a
less intuitive level, similar situations have been studied by Kato
\cite{Kato}. He considered certain finite-matrix toy Hamiltonians
$H(\lambda)$ in complex plane of $\lambda$ and deduced that the
related (in general, complex) spectra $E_n(\lambda)$ change
smoothly with the variation of the parameter $\lambda$ unless one
encounters certain critical, ``exceptional" points
$\lambda^{(EP)}$.

Several recent microwave measurements \cite{Heiss} confirmed the
observability of the abstract Kato's exceptional points
$\lambda^{(EP)}$ in practice.  These experiments re-attracted
attention to the theoretical analyses of the EP horizons, say, in
nuclear physics where many nuclei can, abruptly, lose their
stability \cite{Geyer,Rotter}. The presence of EPs may be also
detected in the random-matrix ensembles with various
interpretations \cite{Berry2} and in optical systems (where EPs
are called ``degeneracies" \cite{Berry}). In classical
magnetohydrodynamics the Kato's exceptional points may even happen
to lie {\em inside} the domain of acceptable parameters,
separating merely the different dynamical regimes of the so called
$\alpha^2-$dynamos~\cite{Uwe}.

The obvious theoretical appeal of the problem of stability may be
perceived as one of the explanations of the recent growth of
popularity of the so called pseudo-Hermitian Hamiltonians in
quantum physics \cite{webpage}. Indeed, one of their distinctive
features is that their spectra are real (i.e., observable) in
parametric domains ${\cal D}$ with, sometimes, very complicated
and strongly Hamiltonian-dependent shape of their EP boundaries
$\p {\cal D}$. For an uninterrupted development of their study it
was very fortunate that a successful semiquantitative description
of the spiked-shaped horizons $\p {\cal D}$ has been found, by
Dorey, Duncan and Tateo \cite{DDT}, for an important class of
analytic (polynomial and power-law, often called ${\cal
PT}-$symmetric) potentials $V(x,\lambda)$ with promising relevance
in quantum field theory \cite{Carl,Mielnik}.

In a few of our own recent studies of EPs in pseudo-Hermitian
Hamiltonians $H$ \cite{Hendrik} - \cite{selfdual} we paid detailed
attention to the possibilities of a deeper geometric understanding
of the structure of the domains ${\cal D}\left (H \right )$ of
quasi-Hermiticity (the name means that the spectrum remains real
for parameters inside ${\cal D}$ -- cf. ref.~\cite{Geyer} for an
older, well written introduction of this concept). After we review
some of the known  results in section \ref{sec2} we shall combine,
in section \ref{sec22}, the methods of algebra (of solvable
equations) and analysis (of elementary curves) in a new approach
to the problem. In this way, the list of results of
ref.~\cite{maximal} (based mainly on computer-assisted symbolic
manipulations) and of ref.~\cite{II} (which used, predominantly,
pertubation-expansion methods) will be complemented by a number of
new non-perturbative items. They will be discussed and summarized
in sections~\ref{sec4} and \ref{par7}.


\section{\label{sec2}Matrix models  }

\subsection{Inspiration: two-dimensional Hilbert space
\label{sec2a}}

The first nontrivial schematic illustration of the current
Schr\"{o}dinger's bound-state problem is provided by the
two-by-two real-matrix model
 \ben
 H\,|\psi\rangle = E\,|\psi\rangle\,,\ \ \ \ \
  H=H(a,b,d)=
 \left (
 \begin{array}{cc}
 a&b\\
 b&d
 \ea
 \right )=H^\dagger(a,b,d)\,.
 \een
Its three-parametric spectrum is {\em always} real and, therefore,
observable,
 \ben
 E=E_\pm(a,b,d) =
 \frac{1}{2}\left [
 a+d\pm \sqrt{(a-d)^2+4\,b^2}
 \right ]\,.
 \een
In the context of ${\cal PT}-$symmetric Quantum Mechanics
\cite{Carl,PTSQM}, the parallel two-by-two example is very similar
 \ben
 H=H'(a,b,d)=
 \left (
 \begin{array}{cc}
 a&b\\
 -b&d
 \ea
 \right )\,, \ \ \ \ E=E'_\pm(a,b,d) =
 \frac{1}{2}\left [
 a+d\pm \sqrt{(a-d)^2-4\,b^2}
 \right ]\,.
 \een
It can {\em still} be considered Hermitian (or, in the language of
the review paper \cite{Geyer}, quasi-Hermitian) {\em after} one
redefines the Hilbert space accordingly (cf. \cite{Hendrik} for
more details).

In the language of phenomenology, one notices an important
complementarity between the parameter-dependence of the two toy
spectra $E_\pm(a,b,d)$ and $E'_\pm(a,b,d)$. In the ``classical",
former example, {\em all} of the energies $E_\pm(a,b,d)$ remain
safely real. The second, primed model is less easy to deal with.
There exists the whole set of the eligible two-by-two metric
operators $\Theta =\Theta^\dagger > 0$ which define the inner
product in the corresponding two-dimensional toy Hilbert space
${\cal H}'$ (cf. \cite{PLB2}). Thus, in spite of its manifest
non-Hermiticity in the auxiliary two-dimensional Hilbert space
${\cal H}$ (where the metric is the Dirac's simplest identity
operator), the operator $H'(a,b,d)$ represents an observable and
remains safely compatible with all the postulates of Quantum
Mechanics (cf. reviews \cite{Geyer,Carl} for more details).

For $H=H(a,b,d)=H^\dagger$ the three-dimensional physical domain
${\cal D}(a,b,d)$ of parameters giving real spectra coincides with
{\em all} $I\!\!R^3$. In contrast, for each individual choice of
the parameters $a$, $b$ and $d$, the quasi-Hermiticity property of
the primed Hamiltonian $H'(a,b,d)$ must be guaranteed and proved.
In general, the reality of the bound-state energies
$E_\pm'(a,b,d)$ and/or the stability of the primed system can only
be achieved inside a {\em perceivably smaller} domain ${\cal
D}'(a,b,d)$ with the easily specified EP horizon,
 \ben
 \p {{\cal D}}'(a,b,d)= \left \{\,
 (a,b,d)\in I\!\!R^3 \ba
 \\
 \ea
 \!\!\right |\, \left .\ba
 \\
 \ea\!\!
 (a-d)^2=4\,b^2
 \right \}\,.
 \een
Thus, the interior of the non-compact manifold ${\cal D}'(a,b,d)$
is specified by the single elementary constraint $ b \in (-|a-d|,
|a-d|)$.

Of course,  for {\em qualitative} considerations the variability
of parameters $a$ and $d$ is entirely redundant. It makes sense to
get rid of them by the multiplicative re-scaling of all the
parameters and by the subsequent shift of the energy scale leading
to the ``generic" choice of $a=-1$ and $d=1$. An extension of this
argument has been formulated in refs.~\cite{PLB3} -- \cite{II}. A
support has been found there for the study of the very special
family of matrix models $H^{(N)}$. In our present paper we just
intend to add new results showing that and how the respective
quasi-Hermiticity domains ${\cal D}$ can be described, at the
lowest dimensions, by non-numerical means.

\subsection{Tridiagonal chain models
\label{sec2b} }

One of the most surprising byproducts of our studies
\cite{Hendrik} - \cite{selfdual} was an empirical observation that
certain particularly simple (i.e., in the sense of regularity,
``canonical") shapes of the $J-$parametric quasi-Hermiticity
domains ${\cal D}^{(N)}$ can be found for a class of the generic,
maximally simplified pseudo-Hermitian Hamiltonians $H^{(N)}$
chosen in the following $N-$dimensional and tridiagonal
``self-dual" \cite{Shifman} matrix form
  \be
 H^{(N)}
 =\left [\begin {array}{cccccc}
  -(N-1)&g_1&&&&\\
 -g_1& -(N-3)&g_{2}&&&\\
 &-g_{2}&\ddots&\ddots&&
 \\
 &&\ddots&N-5&g_{2}&
 \\
 &&&-g_{2}&N-3&g_{1}\\
 &&&&-g_{1}&N-1
 \end {array}\right ]\,
 \label{hamm}
 \ee
with a $J-$plet of real couplings $\vec{\lambda}=(g_1, g_2,
\ldots, g_J)$ and with the dimensions $N=2J$ or $N=2J+1$. In a
more explicit formulation, at any dimension $N$ we found the
coordinates of all the maximal-coupling spikes of the horizon $\p
{\cal D}^{(N)}$ in closed form,
 \be
 g_n^{(spike)}=\pm (N-n)\,n\,,\ \ \ \ \ n = 1, 2, \ldots, J\,.
 \label{spikes}
 \ee
Although this result looks easy, its derivation from the
underlying algebraic equations required extensive
computer-assisted symbolic manipulations and nontrivial
extrapolation guesswork~\cite{maximal}. Moreover, this closed-form
description of the positions of the protruded spikes of the
horizon $\p {\cal D}^{(N)}$ (called, in \cite{maximal},
``extremely exceptional" points, EEPs) has been complemented,  in
our subsequent paper \cite{II}, by the strong-coupling description
of $\p {\cal D}^{(N)}$ based on perturbation ansatz
 \be
 g_n= g_n^{(spike)} \,\sqrt{\left(1-\gamma_n(t) \right) }\,,
 \ \ \ \ \ \ \ \ \
 \gamma_n(t) = t+t^2+\ldots+t^{J-1}+G_n t^J\,.
 \label{lobkov}
  \label{optima}
 \ee
This formula extrapolated, to all $J$, the rigorous $J\leq 2$
fine-tuning rules as derived in refs.~\cite{Hendrik,maximal}. {\it
A posteriori}, using sufficiently small ``redundant" parameters $t
\ll 1$, it proved valid in nonempty open vicinities of all the EEP
vertices.

At the larger $t$s, i.e., far from the EEP spikes, the
determination of the physical horizons $\p {\cal D}^{(N)}$ of our
models $H^{(N)}$ with $J$ free real parameters becomes a more or
less purely numerical task at the higher $J$s \cite{condit}. Up to
now, non-numerical exceptions with $N=2$ and $N=3$ have been
reported in \cite{Hendrik} (where the very easy localization of
the one-parametric interval ${\cal D}^{(2)}\,\equiv\,(-1,1) $ of
$g_1$ has been made) and in \cite{maximal} (mentioning the very
similar result ${\cal D}^{(3)}\,\equiv\,(-\sqrt{2},\sqrt{2}) $).
For the next two dimensions $N=4$ and $N=5$ with two parameters,
the explicit construction of the planar curves $\p {\cal D}^{(N)}$
can be also found in ref.~\cite{maximal}. In what follows we
intend to complement and extend these observations beyond $J=2$
and to show that the closed-form constructions of the prototype
horizons $\p {\cal D}^{(N)}$ remain feasible up to the dimension
as high as $N=11$.

\subsection{Secular equations
\label{sec2bbc} }

Once we choose $N=2J$ or $N=2J+1$, abbreviate $E^2=s$ and, at all
the odd dimensions $N=2J+1$, ignore the persistent and trivial
``middle" energy level $E_J^{(2J+1)}=0$, we find out
\cite{maximal} that all the secular equations $\det \left (
H^{(N)} - E \right )=0$ have the same polynomial form,
 \be
 s^J-\left (
 \ba
 J\\1
 \ea
 \right )\,s^{J-1}\,P +\left (
 \ba
 J\\2
 \ea
 \right )\,s^{J-2}\,Q -\left (
  \ba
  J\\3
  \ea
  \right )\,s^{J-3}\,R +
 \ldots = 0\,.
 \label{polyform}
 \ee
At all $J$ and $N$, the coefficients $P, Q, R, \ldots$ are {real}
polynomial functions of the squares $g_k^2$ of the $J-$plets of
our {real} matrix elements. Once all the energies are assumed real
(i.e., equivalently, once all the roots $s_k$ of
eq.~(\ref{polyform}) happen to be non-negative), we immediately
deduce the following relations tractable as necessary conditions
imposed upon our coefficients in (\ref{polyform}),
 \be
 \ba
 \left (
 \ba
 J\\1
 \ea
 \right )\cdot P
  =
 s_1+s_2+\ldots+s_J\geq 0
  \,,\\
 \left (
 \ba
 J\\2
 \ea
 \right )\cdot Q
  =
 s_1s_2+s_1s_3+\ldots+s_1s_J+s_2s_3+s_2s_4+\ldots +s_{J-1}s_J\geq 0
  \,,\\
 \left (
 \ba
 J\\3
 \ea
 \right )\cdot R
  =
 s_1s_2s_3+s_1s_2s_4+\ldots+s_{J-2}s_{J-1}s_J\geq 0
  \,,\\
 \ldots\ .
  \ea
 \label{glyptoform}
 \ee
In the opposite direction, the set of the necessary inequalities
$P \geq 0$, $Q \geq 0$, $\ldots$ is incomplete as it does not
provide the desirable sufficient condition of quasi-Hermiticity.
It admits complex roots $s$ in general (take a sample secular
polynomial $(s^2+1)(s-2)$ for illustration).

\section{Domains ${\cal D}^{(2J)}$ and ${\cal D}^{(2J+1)}$
\label{sec22}}

Obviously, for a given prototype Hamiltonian $H^{(N)}$ and under
the constraints (\ref{glyptoform}), the determination of the
quasi-Hermiticity domain ${\cal D}^{(N)} = {\cal D}\left (
H^{(N)}\right )$ is {\em equivalent} to the guarantee of the
non-negativity of all the $J$ roots $s_k$  of
eq.~(\ref{polyform}). The explicit forms of the corresponding
sufficient conditions will now be given for the first ten smallest
matrix dimensions $N= 2, 3, \ldots, 11$.

\subsection{ Non-negativity of the root of
eq.~(\ref{polyform}) at $J=1$  \label{sec2ba}}

At $J=1$ the linear version $ s-P=0$ of secular
eq.~(\ref{polyform}) has the single root $s_0 = P$. The
non-negativity of this root is equivalent to the non-negativity of
the coefficient $P$. This means that in terms of the single
coupling $g_1=a$ available at $J=1$, the necessary and sufficient
criteria of the observability of $H^{(2)}=H^{(2)}(a)$ or
$H^{(3)}=H^{(3)}(a)$ read $P^{(2)}(a)=1-a^2\geq 0$ and
$P^{(3)}(a)=4-2\,a^2\geq 0$, respectively. Thus, in a way
transferable, {\it mutatis mutandis}, to any dimension, the
explicit definitions ${\cal D}^{(2)}(a) = (-1,1)$ and ${\cal
D}^{(3)}(a) = (-\sqrt{2},\sqrt{2})$ of the quasi-Hermiticity
domains may be re-read as definitions of the corresponding EP
horizons $\p {\cal D}^{(2)}(a) = \{-1,1\}$ and $\p {\cal
D}^{(3)}(a) =\{-\sqrt{2},\sqrt{2}\}$.

\subsection{Non-negativity of all the
roots of eq.~(\ref{polyform}) at $J=2$ \label{sec2c} }

At $J=2$ the quadratic version $ s^2-2\,P\,s^{} +Q=0$ of secular
eq.~(\ref{polyform}) has two roots $s_\pm = P\pm \sqrt{P^2-Q}$.
These two roots remain real if and only if $B \equiv P^2-Q \geq
0$. In the subdomain of parameters where $B \geq 0$ they remain
both non-negative if and only if $P\geq 0$ and $Q \geq 0$. We can
summarize that the required sufficient criterion reads
 \be
 P \geq 0\,,\ \ \ \
 P^2 \geq Q \geq 0\,.
 \label{requi}
 \ee
In an alternative approach, {\em without} an explicit reference to
the available formula for $s_\pm$, let us contemplate the
parabolic curve $y(s)=s^2-2\,P\,s$ which remains safely positive,
in the light of our assumption (\ref{glyptoform}), at all the
negative $s<0$. This means that this curve can only intersect the
horizontal line $z(s)=-Q$ at some non-negative points $s\geq 0$.

In this way the proof of non-negativity of all the roots of our
secular equation degenerates to the proof that there exist two
real points of intersection of the $J=2$ parabola $y(s)$ with the
horizontal line $z(s)$ (which lies below zero) at some $s\geq 0$.
Towards this end we consider the minimum of the curve $y(s)$ which
lies at the point $s_0$ such that $y'(s_0)=0$, i.e., at $s_0=P$.
This minimum must lie {\em below} (or, at worst, at) the
horizontal line of $z(s)=-Q\leq 0$. But the minimum value of
$y(s_0)$ is known, $y(P)=-P^2$. Thus, the condition of
intersection $y(s_0) \leq z(s_0)$ gives the formula $P^2 \geq Q$.
QED.

Marginally, it is amusing to notice that once
eq.~(\ref{glyptoform}) holds, the inequality $P^2-Q\geq 0$ is
equivalent to the reality of the roots simply because
$P^2-Q\,\equiv\,(s_1-s_2)^2/4$. In fact, even for some other
two-parametric matrices, precisely {this} type of requirement is
responsible for an important part of the EP boundary $\p {\cal D}$
(cf. refs.~\cite{PLB3,determ} for details).

\subsection{Non-negativity of all the
roots of eq.~(\ref{polyform}) at $J=3$ \label{sec2d} }

Neither at $N=6$ nor at $N=7$ the sufficient condition of
non-negativity of all the energy roots $s$ is provided by the
three necessary rules $P\geq 0$, $Q\geq 0$ and $R \geq 0$ of
eq.~(\ref{glyptoform}). Let us return, therefore, to the second
method used in paragraph \ref{sec2c} and derive another,
``missing" inequality needed as a guarantee of the reality of the
energies. In the first step one notices again that all the three
components of the polynomial
 \ben
 y(s)=s^3-3\,P\,s^2+3\,Q\,s=R\,, \ \ \ \ \ \ \ J=3
 \een
remain safely non-positive at $s<0$. Whenever the roots are
guaranteed real, their non-negativity  $s_n\geq 0$ with $n=1,2,3$
is already a consequence of the three constraints
(\ref{glyptoform}). The necessary condition of their reality is
less trivial but it still can be deduced from the shape of the
function $y(s)$ on the half-axis $s\geq 0$, i.e., from the
existence and properties of a real maximum of $y(s)$ (at $s=s_-$)
and of its subsequent minimum (at $s=s_+$). At both these points
the derivative $y'(s)=3\,s^2-6\,P\,s+3\,Q$ vanishes so that both
the roots $ s_\pm = P \pm \sqrt{P^2-Q}$ of $y'(s)$ must be real
and non-negative. This condition is always satisfied for the {\em
real} roots $s_k$ of $y(s)$ since
 \ben
 B= P^2-Q\ \equiv\ \frac{1}{54}\,
 \left [
 \left (
 s_1+s_2-2\,s_3
 \right )^2+
 \left (
 s_2+s_3-2\,s_1
 \right )^2+
 \left (
 s_3+s_1-2\,s_2
 \right )^2
 \right ]\geq 0\,.
 \een
In the next step, the necessary guarantee of the reality of the
roots $s_k$ will be understood again as equivalent to the doublet
of the inequalities $y(s_-)\geq R$ and $y(s_+)\leq R$. Here we may
insert $s_\pm^2=2Ps_\pm-Q$ and get the two inequalities which are
more explicit,
 \be
 2(P^2-Q)\,s_-\ \leq \ PQ-R\ \leq \
 2(P^2-Q)\,s_+\,.
 \label{uhrad}
 \ee
They restrict the range of a new symmetric function of the roots,
 \ben
 P\,Q-R\ \equiv\ \frac{1}{9}\,
 \left [s_1s_2
 \left (
 s_1+s_2-2\,s_3
 \right )+s_2s_3
 \left (
 s_2+s_3-2\,s_1
 \right )+s_3s_1
 \left (
 s_3+s_1-2\,s_2
 \right )
 \right ]\,.
 \een
After another insertion of the known $s_\pm$ we arrive at a
particularly compact formula
 \ben
  2(P^2-Q)^{3/2} \geq R-3\,PQ+ 2P^3 \geq
 -2\,(P^2-Q)^{3/2}\,
 \een
or, equivalently,
 \ben
  4\,\left (P^2-Q \right )^{3} \geq
  \left (R-3\,PQ+ 2P^3\right )^2\,.
 \label{uhradu}
 \een
Due to the numerous cancellations the latter relation further
degenerates to the most compact {missing necessary condition}
 \be
  3\,{P}^{2}{Q}^{2}+6\,RPQ
  \geq 4\,{Q}^{3}+{R}^{2}+4\,R{P}^{3}\,.
 \label{uhrada}
  \ee
Our task is completed. In combination with
eqs.~(\ref{glyptoform}), equation (\ref{uhrada}) plays the role of
the guarantee of the reality of the energy spectrum.

\subsection{Non-negativity of all the
roots of eq.~(\ref{polyform}) at $J=4$ \label{sec2e} }

In a search for the non-negative roots of the quartic secular
equation
 \be
 \det\left ( H^{(8,9)}-E\,I\right ) =
 x^4-4\,P\,x^3+6\,Q\,x^2-4\,R\,x+S\,\equiv\,y(x)+S=0
 \,
 \label{jejichkuba}
 \ee
we note that all the four $N-$dependent coefficients $P$, $Q$, $R$
and $S$ again evaluate as certain polynomials in the squares of
the four coupling parameters $g_k$, $k = 1,2,3,4$. Once all these
four expressions are kept non-negative, the curves $y(x)$ and
$z(x)=-S$ do not intersect at $x<0$. At $x \geq 0$ they do
intersect four times at $x\geq 0$ (as required), provided only
that the three extremes of $y(x)$ can be found at the three
non-negative real roots $x_{1,2,3}$ of the extremes-determining
equation
 \be
 y'(x_{1,2,3})=4\,(x_{1,2,3}^3-3\,P\,x_{1,2,3}^2
 +3\,Q\,x_{1,2,3}-R)=0\,.
 \ee
In an ordering $0 \leq x_1\leq x_2\leq x_3$ of these roots we
arrive at the three sufficient conditions
 \be
 y(x_1)\leq -S\,,\ \ \ \ \
 y(x_2)\geq -S\,,\ \ \ \ \
 y(x_3)\leq -S\,
 \label{jeknub}
 \ee
guaranteeing that the parameters lie inside ${\cal D}^{(8)}$ or
${\cal D}^{(9)}$.

All the three quantities $x_k$ satisfy the cubic equation
$y'(x)=0$ so that its premultiplication by $x$ enables us to
eliminate the three fourth powers from $y(x)$,
 \ben
 x_{1,2,3}^4=3\,P\,x_{1,2,3}^3-3\,Q\,x_{1,2,3}^2+R\,x_{1,2,3}\,.
 \een
Their insertion reduces all the three items of eq.~(\ref{jeknub})
to the other three intermediate polynomial inequalities of the
third degree,
 \ben
  -P\,x_{1,3}^3+3\,Q\,x_{1,3}^2-3\,R\,x_{1,3}+S \leq 0\,,
  \label{ajejichneknuba1}
 \een
 \ben
  -P\,x_{2}^3+3\,Q\,x_{2}^2-3\,R\,x_{2}+S \geq 0\,.
  \label{ajejichneknuba2}
 \een
Repeating the same trick once more, the elimination of
 $
 x_{1,2,3}^3=3\,P\,x_{1,2,3}^2-3\,Q\,x_{1,2,3}+R
 $
gives an equivalent triplet of inequalities
 \begin{eqnarray}
  -
  B\,x_{1}^2
 +2\,B^{3/2}\,C
  \,x_{1} \leq B^2\,D\,,
  \label{bcjejichneknuba2}
 \\
  -
  \label{bcdjejichneknuba2}
  B\,x_{2}^2
 +2\,B^{3/2}\,C
  \,x_{2} \geq B^2\,D\,,\\
  -
  B\,x_{3}^2
 +2\,B^{3/2}\,C
  \,x_{3} \leq B^2\,D\,.
  \label{bjejichneknuba2}
 \end{eqnarray}
Here, the old abbreviations $B=P^2-Q$ and $2\,B^{3/2}\,C=PQ-R$
plus a new one, $3\,B^2\,D=P\,R-S $ enable us to re-scale
$x_{1,2,3,}=\sqrt{B}\,Y_{1,2,3}$ which yields our final triplet of
quadratic-equation conditions
 \begin{eqnarray}
  Y_{1}^2
 -2\,C
  \,Y_{1} +D\geq 0\,,
  \label{xbcjejichneknuba2}
 \\
    \label{xbcdjejichneknuba2}
  Y_{2}^2
 -2\,C
  \,Y_{2}+D \leq 0\,,\\
  Y_{3}^2
 -2\,C
  \,Y_{3}+D \geq 0\,.
  \label{xbjejichneknuba2}
 \end{eqnarray}
The auxiliary roots $Y_\pm = C \pm \sqrt{C^2-D}$ must be real and
non-negative. In this way we must guarantee that $D\geq 0$ and
$C^2\geq D$. The conclusion is that eqs.~(\ref{xbcjejichneknuba2})
-- (\ref{xbjejichneknuba2}) degenerate to the four final
elementary requirements
 \be
 Y_1\leq Y_-\leq Y_2\leq Y_+\leq
 Y_3\,.
 \label{finalistj}
 \ee
They complement the inequalities  $B\geq 0$, $Q\geq 0$ and $-1
\leq C-\sqrt{1+Q/B}\leq 1$ and form the complete algebraic
definition of the domains ${\cal D}^{(8)}$ and ${\cal D}^{(9)}$.

\subsection{Non-negativity of all the
roots of eq.~(\ref{polyform}) at $J=5$ \label{sec2f} }

Let us finally proceed to  $H^{(N)}$ with $N=10$ and/or $N=11$
which leads to the ``unsolvable" secular equations of the fifth
degree,
 \be
 x^5-5\,P\,x^4+10\,Q\,x^3-10\,R\,x^2+5\,S\,x-T\,\equiv\,y(x)-T=0
 \,.
 \label{zzjejichkuba}
 \ee
From our present point of view the problem of the construction of
the respective horizons $\p {\cal D}^{(N)}$  remains solvable
exactly since the derivative $y'(x)$ is still of the mere fourth
degree,
 \be
 \frac{1}{5}\,y'(x)=
 x^4-4\,P\,x^3+6\,Q\,x^2-4\,R\,x+S \,.
 \,
 \label{zzjuba}
 \ee
The exact, real and non-negative values $x_{1}\leq x_2\leq x_3\leq
x_4$ of the four roots of $y'(x)$ may still be considered
available in closed form.

In a way which parallels our preceding considerations we may
assume that the five $N-$dependent non-negative coefficients
$P\geq 0$, $Q\geq 0$, $R\geq 0$, $S\geq 0$ and $T\geq 0$ obey also
all the additional inequalities derived in the preceding sections.
In a more detailed description, we may then treat our secular
problem (\ref{zzjejichkuba}) as a search for the graphical
intersections between the (nonnegative) constant curve $z(x)=T$
and the graph of the polynomial $y(x)$ of the fifth degree (which
can only be nonnegative at $x\geq 0$). Inside the domain ${\cal
D}^{(N)}$, the quintuplet of the (unknown but real and
nonnegative) physical energy roots $x_a$, $x_b$, $x_c$, $x_d$ and
$x_e$ may be assumed compatible with the obvious intertwining rule
 \ben
 0\leq x_a\leq x_1\leq x_b
 \leq x_2\leq x_c\leq x_3
 \leq x_d\leq x_4\leq x_e\,.
 \een
The way towards the sufficient condition of the existence of the
real energy spectrum remains the same as above, requiring
 \be
 y(x_1)\geq T\,,\ \ \ \ \
 y(x_2)\leq T\,,\ \ \ \ \
 y(x_3)\geq T\,,\ \ \ \ \
 y(x_4)\leq T\,.
 \label{zzjeknub}
 \ee
The lowering of the degree should again reduce
eq.~(\ref{zzjeknub}) to the quadruplet
 \be
 w(Y_1)\leq 0\,,\ \ \ \ \
 w(Y_2)\geq 0\,,\ \ \ \ \
 w(Y_3)\leq 0\,,\ \ \ \ \
 w(Y_4)\geq 0\,.
 \label{wwzzjeknub}
 \ee
where the re-scaling $x_{1,2,3,4}=Y_{1,2,3,4}\,\sqrt{B}$ applies
to the arguments of the brand new auxiliary polynomial function of
the third degree in $Y$,
 \ben
 w(Y)=Y^3-3\,C\,Y^2+3\,D\,Y-G\,.
 \een
Besides the same abbreviations as above, we introduced here a new
one, for $PS-T\,\equiv\,4\,B^{5/2}\,G$. The new and specific
problem now arises in connection with the necessity of finding the
three auxiliary and, of course, real and non-negative roots of the
cubic polynomial $w(Y)$. Once we mark them, in the ascending
order,  by the Greek-alphabet subscripts, we should either
postulate our (in principle, explicit) knowledge of their real and
nonnegative values $Y_\alpha\leq Y_\beta\leq Y_\gamma$ or, in
another perspective, we have to add the above-studied conditions
which restrict the range of the three coefficients $C$, $D$ and
$G$ in the cubic polynomial $w(Y)$.


This being said, the rest of the story is easy to tell. Once we
parallel the same geometric argument as used in our previous
subsections, we may immediately conclude that our ``last feasible"
specification of the domains ${\cal D}^{(10)}$ and ${\cal
D}^{(11)}$ will be given by the following set of the inequalities,
 \be
 Y_1\leq Y_\alpha\leq Y_2\leq Y_\beta\leq
 Y_3\leq Y_\gamma\leq
 Y_4\,.
 \label{wwfinalistj}
 \ee
This is the last algebraic formula which defines the domains
${\cal D}^{(10)}$ and ${\cal D}^{(11)}$. Any extension of the
recipe beyond $N =11$ would suffer from the necessity of using
certain purely numerically defined auxiliary functions of
couplings $g_k$.

\section{Discussion
\label{sec4} }

\subsection{Reparametrizing the couplings
 \label{sec3} }

The existence of the algebraic formulae which determine all the
boundaries $\p {\cal D}^{(N)}$ up to $N=11$ opens a way towards a
non-perturbative extension of the results of refs.~\cite{maximal}
and \cite{II} beyond the strong-coupling dynamical regime. In such
a setting, the old perturbation ansatz (\ref{lobkov}) can be
re-interpreted as a {\em precise}, non-perturbative change of
variables $g_k \ \longrightarrow \ G_k$. During such a process,
the redundant value of $t$ may be fixed arbitrarily (cf. the
construction of the planar curve $\p {\cal D}^{(4)}$ in section
3.1 of paper \cite{II} for illustration). Due to the reflection
symmetries of our hedge-hog-shaped horizons $\p {\cal D}^{(N)}$,
one is also allowed, without any loss of generality, to restrict
attention to the subdomain of $ {\cal D}^{(N)}$ with positive
$g_k$s. In such a setting the new, rescaled real couplings
$\gamma^{(N)}_J=\alpha$, $\gamma^{(N)}_{J-1}=\beta$, $\ldots$
should remain non-negative and smaller than one,
$\gamma^{(N)}_k\in (0,1)$. For illustration one may recollect the
most elementary two-by-two Hamiltonian
 \be
 H^{(2)}= \left (
 \begin{array}{cc}
 -1&\sqrt{1-\alpha}\\
 -\sqrt{1-\alpha}&1
 \ea
 \right )\,,\ \ \ \ \alpha \in (0,1)\,
 \label{uno}
 \ee
with the two-point spectrum $E_\pm^{(2)} = \pm \sqrt{ \alpha} \,$.
With respect to the new parameter we have to set ${\cal D}^{(2)}
(\alpha)\ \equiv\ (0,1)$ since there are no additional
constraints.

\subsection{New forms of approximations
 \label{sec3aaf} }

We should emphasize that in comparison with our previous work, the
most significant progress has been achieved here in the new
non-approximate and complete description of the structure of the
boundaries of the domains ${\cal D}^{(N)}$ at $N=6$ and $N=7$ by
means of our key  inequality (\ref{uhrad}). In order to stress
some of its merits, let us now add a few comments on this form of
the rigorous guarantee of the reality of the energies at $J=3$.

Firstly, let us set $P^2=B+Q$ and switch just to the postulates $B
\geq 0$ and $Q \geq 0$, re-classifying the expression
$P=+\sqrt{B+Q}$ itself as a mere formal abbreviation. This enables
us to insert
 \ben
 s_\pm = \sqrt{B+Q} \pm \sqrt{B} \geq 0\,
 \een
in eq.~(\ref{uhrad}) which prescribes $s_-\ \leq \ C\,{\sqrt{B}}
\leq \ s_+$ with $PQ-R=2\,B^{3/2}\,C\geq 0$, i.e.,
 \be
 {\sqrt{1+q} - 1}\ \leq \ {C}\ \leq \
 {\sqrt{1+q} + 1} \,, \ \ \ \ \ \ \  q = \frac{Q}{B}\,\in \,(0,\infty)\,.
  \label{uhradeb}
 \ee
Once we notice that
$PQ\,\equiv\,Q\,\sqrt{B+Q}=q\,B^{3/2}\sqrt{1+q}$ we may return
from the auxiliary $C$ to the original $R$ and rewrite
eq.~(\ref{uhradeb}) as a perceivably simplified one-parametric
constraint
 \be
 1+\left ( \frac{q}{2}-1\right )
 \sqrt{1+q}
 \geq
 \frac{R}{2\,B^{3/2}}
 \geq
 \left \{
 \begin{array}{ll}
 0,&q\leq 3\,,\\
 \left ( \frac{q}{2}-1\right )
 \sqrt{1+q}\,-1\,,&q>3\,
 \ea
 \right .
 \label{twosi}
 \ee
imposed upon the rescaled polynomial $R$. It completes our
specification of the physical domain ${\cal D}^{(6,7)}$ in terms
of the coefficients $B\geq 0$, $Q\geq 0$ and $R$. We see that with
 \ben
 1+\left ( \frac{q}{2}-1\right )
 \sqrt{1+q}=
{\frac {3}{8}}{q}^{2}-{\frac {1}{8}}{q}^{3}+{\frac
{9}{128}}{q}^{4}-{ \frac {3}{64}}{q}^{5}+{\frac
{35}{1024}}{q}^{6}+{\cal O}\left ({q}^{7}\right )\,,
 \een
the two-sided inequality (\ref{twosi}) is fairly restrictive in
the $R-$direction, especially at the smallest ratios $q=Q/B$.

\subsection{Pairwise
confluences of the levels \label{par3.2} }

Several aspects of the ``first nontrivial" $J=3$ problem have been
skipped in the main text but they definitely deserve more
attention in the discussion. For the sake of definiteness let us
choose just $N=6$ and use
 \ben
 g_1=c=\sqrt{5\,(1-\gamma)}\,,\ \ \ g_2=b= 2\,\sqrt{2\,(1-\beta)}\,,
 \ \ \ g_3=a= 3\,\sqrt{1-\alpha}
 \een
with parameters $\alpha,\beta,\gamma \in (0,1)$ entering the
six-by-six matrix (\ref{hamm}). Its secular equation
(\ref{polyform}) determines the spectrum which remains real, by
definition, whenever all the couplings lie inside the domain
${\cal D}^{(6)}$. It is easy to deduce that the latter domain is
circumscribed by the ellipsoidal surface given by the equation
 \ben
 P=-\left ({a}^{2}+2\,{b}^{2}+2\,{c}^{2}-35\right )/3=0\,.
 \een
The other two obvious constraints read
 \ben
 3\,Q=
 {b}^{4}+2\,{c}^{2}{a}^{2}-44\,{b}^{2}
 +28\,{c}^{2}-34\,{a}^{2}+{c}^{4}+
 259+2\,{b}^{2}{c}^{2}\geq 0
 \een
and
 \ben
 -R=
 {a}^{2}{c}^{4}-10\,{b}^{2}{c}^{2}
 +30\,{c}^{2}{a}^{2}+225\,{a}^{2}-30\,{c}^{2}
 -{c}^{4}-25\,{b}^{4}-225-150\,{b}^{2}\geq 0\,.
 \een
The last constraint needed to define ${\cal D}^{(6)}$ is then
given by eq.~(\ref{uhrad}). In its light one can spot certain new
structures in $\p {\cal D}^{(6)}$ where, for example, the  total
EEP confluence of all the energies can be preceded by some
incomplete, pairwise coincidences among the six levels in
question. For example, an ``innermost" pair of the energies can
coincide at $E=0$ while, independently, the other two ``outer"
doublets of the energies are allowed to coincide at the two
symmetric non-vanishing values $E=\pm 4z$ at an unknown parameter
$z\in (0,1)$. Alternatively, the two ``outermost" levels [with
$s=s_{max}=5\,y$ where $y \in (0,1)$] can stay real while the
confluence only involves the two internal energy doublets at a
shared value of $s=s_{min}=4\,x$ where $x \in (0,1)$.

In the latter scenario one has to reproduce the two-parametric
relation
 \ben
 (s-s_{max})\,\left [s-s_{min} \right ]^2=
 s^3-
 \left (32\,x^2+25\,y^2 \right )\,s^2
 +\left (
 256\,x^4+800\,x^2y^2
 \right )\,s
 -6400\,x^4y^2=0\,
 \een
which, implicitly, defines the third sub-surface. Let us
concentrate on the former, slightly simpler scenario where the
surface of the three-dimensional domain ${\cal D}^{(6)}$ can be
visualized as composed, locally, of the two eligible smooth
sub-surfaces which intersect along a certain ``double exceptional
point" (DEP) curve.

\subsection{Pairwise confluences of exceptional points}

In terms of the single free parameter $z$ of the latter particular
scenario, the DEP secular equation degenerates to the formula $
 E^2\,\left (E+4z \right )^2\,\left (E-4z \right )^2=0
$, to be obtained, in the one-parametric DEP limit, from
eq.~(\ref{polyform}), i.e.,
 \be
 s\,\left [s-(4z)^2 \right ]^2=
 s^3-2\,(4z)^2s^2+(4z)^4 s=0\,.
 \label{depka}
 \ee
It is fortunate that the necessary analysis can still be performed
non-numerically since equation~(\ref{depka}) is easy to compare
with the true secular equation (\ref{polyform}) with coefficients
given in paragraph \ref{par3.2}.
%
As long as the factorizable coefficient at $s^0$ must vanish, we
get the first DEP constraint
 $$ \left [a{c}^{2}+15\,a
 + \left (15+{c}^{2}+5\,{b}^{2}\right )\right ]\,
  \left [a{c}^{2}+15\,a
 -\left (15+{c}^{2}+5\,{b}^{2}\right )\right ]
 =0 \,$$
so that we may eliminate
 \ben
 a=\pm \frac{15+{c}^{2}+5\,{b}^{2}}{{c}^{2}+15}\,.
 \een
In the quadrant of $a-b-c$ space with positive $a$ the plus sign
must be chosen,
 \ben
 a= 1 +\frac{5\,{b}^{2}}{{c}^{2}+15} \,
 \een
i.e., we have $3 \geq a \geq 1$ in the closed formula for
 \be
 b^2=\frac{1}{5}\,({c}^{2}+15)\,(a -1)\,
 \label{fei}
 \ee
or, alternatively, for
 \ben
 c^2=\frac{5b^2}{a-1}-15\,.
 \een
This result is to be complemented by the other two relations
  \ben
 3\,Q(c,b,a)=32\,z^2\,,\ \ \ \ \ \ \
 R(c,b,a)=128\,z^4\,.
 \een
A straightforward elimination of $z^2$ gives the second DEP
condition
 \ben
  -66\,{a}^{2}-36\,{b}^{2}+4\,{c}^{2}{a}^{2}-189
  +252\,{c}^{2}-4\,{b}^{2}{a}^{2}-{a}^{4}=0
 \een
with the two compact roots
 \ben
 a^2_{\pm}=
 2\,{c}^{2}-33-2\,{b}^{2}\pm
 2\,\sqrt {{c}^{4}+30\,{c}^{2}-2\,{b}^{2}{c}^{2}
 +225+24\,{b}^{2}+{b}^{4}}\,.
 \een
The acceptable one must be non-negative. For $a^2_-$ this would
mean that $2\,{c}^{2}\geq 33+2\,{b}^{2}$ while, at the same time,
$
 63+12\,{b}^{2}\geq 84\,{c}^{2}
$. These two conditions are manifestly incompatible so that we
must accept the upper-sign root $a^2_+$ which is automatically
positive for all the large $2\,{c}^{2}\geq 33+2\,{b}^{2}$ and
which remains positive for all the smaller $c^2$ constrained by
the requirement
 \ben
 84\,{c}^{2} \geq 63+12\,{b}^{2}\,.
 \een
After the insertion of the definition (\ref{fei}) of $b^2$ we
arrive at the condition
 \be
 84\,{c}^{2} \geq
 63+{\frac {12}{5}}\,\left (a-1\right )\left
 (15+{c}^{2} \right )\,
 \label{unequal}
 \ee
with a non-empty domain of validity.

\section{Conclusions \label{par7}}


According to the abstract principles of Quantum Mechanics,
observable quantities (say, the energies $E_0<E_1<\ldots$ of bound
states) should be constructed as eigenvalues of certain
self-adjoint operators of observables (i.e., in our illustration,
of a Hamiltonian $H$ acting in some physical Hilbert space of
states ${\cal H}$). Of course, an explicit representation of
${\cal H}$ can prove complicated. Hence, the idea emerged  of an
introduction of a simpler, ``auxiliary" Hilbert space (to be
denoted here as ${\cal H}^{(aux)}$).

Although the origin of the latter idea can be traced back to the
very early days of Quantum Theory \cite{Mielnik}, the feasibility
of its separate implementations have long been treated as a mere
mathematical curiosity (cf., e.g., \cite{BG} for illustration).
Among the people who felt forced to take it really very seriously
were the nuclear-physics specialists. Incidentally, in their so
called IBM models of atomic nuclei, one of the ``natural" (often
called Dyson's) fermion-to-boson mappings ${\cal H}
\leftrightarrow {\cal H}^{(aux)}$ happened to simplify the
solution of the Schr\"{o}dinger equation considerably. In 1992 the
topic has nicely been reviewed by Scholtz et al \cite{Geyer}.

In 1998, many other physicists of different professional
orientations (ranging from supersymmetry \cite{Cannata} and field
theory \cite{BM} to cosmology \cite{Alifirst} etc) have got
involved when Bender and Boettcher \cite{BB} attracted their
attention  to the specific class of examples
 \be
 -\frac{\hbar^2}{2m}\,\frac{d^2}{dx^2}\,\psi^{(aux)}(x)
 + V^{(aux)}(x)\,\psi^{(aux)}(x) = E\,\psi^{(aux)}(x)\,
 \label{SE}
 \ee
(cf. also refs.~\cite{BBb,solvable}) where the potentials may be
complex but where the spectra remain real \cite{DDT}. Thus,
although the physical space ${\cal H}$ itself proves complicated
(mainly, due to its highly nontrivial definition of the inner
product \cite{BBJ}), the auxiliary space ${\cal H}^{(aux)}$ can
often be chosen in its most common representation
$I\!\!L_2(I\!\!R)$ of the square-integrable complex functions of
one variable. In this setting (cf. its recent review \cite{Carl}),
the difference between the clarity of the properties of the space
of functions ${\cal H}^{(aux)}=I\!\!L_2(I\!\!R)$ and the
perceivably more complicated character of the physical states in
${\cal H}$ emerges due to the nontriviality of the ``physical
metric" $\Theta\neq I$ in ${\cal H}$ (cf. ref.~\cite{Geyer} for a
compact review of the necessary mathematical properties of this
operator). In the context of eq.~(\ref{SE}) (possessing also
numerous exactly solvable special cases \cite{solvable}), the
metric $\Theta$ is assumed constructed as the product of certain
operators ${\cal P}$ (= parity) and ${\cal C}$ (= ``charge").

Beyond the specific class of the differential phenomenological
models (\ref{SE}), another obvious choice of ${\cal H}^{(aux)}$
would be finite-dimensional \cite{Geyer,turek}. One of the key
merits of such an alternative extension of the class of the
``tractable" models of bound states lies in the fact that the key
mathematical proofs of the reality of their spectra may become
decisively simpler. It is in the latter context, with $\dim \,
{\cal H}^{(aux)}=N<\infty$, where our attention has been attracted
to the various specific models where one can extract more
information about the {\em shape} of the spectral-reality domain
${\cal D}$ of variable parameters in the Hamiltonians.


As we emphasized in \cite{condit}, a nontrivial relationship may
exist between the matrix, finite-dimensional models and the so
called quantum catastrophes interpreted as changes of some
parameters $\lambda_j$, $j = 1, 2, \ldots, J$ in the Hamiltonians
$H(\vec{\lambda})$ which lead to the loss of the reality (i.e., of
the observability) of certain energies $E_{n_c}(\vec{\lambda})$ in
the spectrum. In a step towards a typology of such a phenomenon we
showed here that for our ${\cal PT}-$symmetric chain models
$H^{(N)}$ of dimensions $N = 2, 3, \ldots, 11$ a non-numerical
knowledge becomes available about the parametric domains ${\cal
D}^{(N)}$ in which all the energies remain real. This means that
the ``admissible" $J-$plets of the coupling constants have been
shown defined by certain non-numerical means. In this way, the
explicit control of the stability of the system becomes mediated
via purely algebraic constraints imposed upon the controllable
parameters.

In the conclusion we feel encouraged to express our belief that
{\em many} qualitative and geometric features of the observability
horizons $\p {\cal D}$ assigned to {\em any given} ${\cal
PT}-$symmetric Hamiltonian $H$ may be expected to {\em survive}
its reduction to a series of $N$ by $N$ approximations of the
prototype form $H^{(N)}$. Such a feature would really enhance the
relevance of our present study of the peculiar self-dual ${\cal
PT}-$symmetric Hamiltonians $H^{(N)}\neq \left ( H^{(N)}\right
)^\dagger$. In particular, it would also assign a deeper meaning
to our present detailed analysis of the boundaries $\p {\cal
D}^{(N)}$ at the first few dimensions  up to $N=11$. In such a
setting and perspective it is also appropriate to re-emphasize the
two reasons of the relevance of our knowledge of the physical
domains $ {\cal D}$. Firstly, their boundaries $\p {\cal D}$ are
marking the breakdown of the reality and observability of the
spectrum $\{E_n\}$. Secondly, all the vicinity of these boundaries
also represents a region where the matrices $H^{(N)}$ cease to be
diagonalizable. Thus, it is the {\em simultaneous} degeneracy of
the energies and of the wave functions which gives the full and
deep physical meaning to this horizon of the dynamical stability
of the system.

\vspace{5mm}

\section*{Acknowledgement}

Work supported by the GA\v{C}R grant Nr. 202/07/1307, by the
M\v{S}MT ``Doppler Institute" project Nr. LC06002 and by the
Institutional Research Plan AV0Z10480505.


%


%
%



\begin{thebibliography}{00}


\bibitem{Landau}
L. D. Landau and E. M. Lifshitz, Quantum Mechanics (Pergamon,
Oxford, 1977).

\bibitem{Greiner}
W. Greiner, Relativistic Quantum Mechanics - Wave Equations
(Springer, Berlin, 1997).

\bibitem{Kato}
 T. Kato, {Perturbation Theory for linear Operators}
(Springer, Berlin, 1966), p. 64.

\bibitem{Heiss}
C. Dembowski et al, Phys. Rev. Lett. 90 (2003) 034101;

D. Heiss, Czech. J. Phys. 54 (2004) 1091.

\bibitem{Geyer}
 F. G. Scholtz, H. B. Geyer and F. J. W.
Hahne, Ann. Phys.
(NY) 213 (1992) 74.

\bibitem{Rotter}
H. B. Geyer, Czech. J. Phys.  54 (2004) 51;

M. S. Swanson, J. Math. Phys. 45 (2004) 585;

I. Rotter, Czech. J. Phys. 55 (2005) 1167;

B. R. Barrett, I. Stetcu, P. Navratil and J. P. Vary, J. Phys. A:
Math. Gen. 39 (2006) 9983.

\bibitem{Berry2}
Z. Ahmed and S. R. Jain, Phys. Rev. E 67 (2003) 045106(R);

S. R. Jain, Czech. J. Phys. 56 (2006) 1021;

M. V. Berry, J. Phys. A: Math. Gen. 39 (2006) 10013.

\bibitem{Berry}
M. V. Berry, Czech. J. Phys. 54 (2004) 1039.

\bibitem{Uwe}
U. G\"{u}nther and F. Stefani, J. Math. Phys. 44 (2003) 3097;

U. Guenther, F. Stefani and M. Znojil,
J. Math. Phys. 46  (2005) 063504;

O. Kirillov and U. G\"{u}nther, J. Phys. A: Math. Gen. 39 (2006)
10057.
%

\bibitem{webpage}
http://gemma.ujf.cas.cz/$\sim $znojil/conf/index.html;

H. Geyer, D. Heiss and M. Znojil (editors),
J. Phys. A: Math. Gen. 39 (2006), Nr. 32 (dedicated special
issue);

M. Znojil (editor), Czechosl. J. Phys. 56 (2006), Nr. 9 (dedicated
special issue).

\bibitem{DDT}
P. Dorey, C. Dunning and R. Tateo, J. Phys. A: Math. Gen. 34
(2001) 5679 and Czech. J. Phys. 54 (2004) 35.

\bibitem{Carl}
C. M. Bender, Reports on Progress in Physics 70 (2007) 947.

\bibitem{Mielnik}
W. Pauli, Rev. Mod. Phys. 15 (1943) 175;

E. C. G. Sudarshan, Phys. Rev. 123 (1961) 2183;

K. L.  Nagy, State Vector Spaces with Indefinite Metric in Quantum
Field Theory (Budapest, Akademiai Kiado, 1966);

T. D. Lee and G. C. Wick, Nucl. Phys. B 9 ((1969) 209;

N. Nakanishi, Phys. Rev. D 3 (1971) 811;

F. Kleefeld, AIP Conf. Proc. 660 (2003) 325;

A. Ramirez and B. Mielnik, Rev. Mex. Fis. 49 S2 (2003) 130.

\bibitem{Hendrik}
M. Znojil and H. B. Geyer, Phys. Lett. B 640 (2006) 52.

\bibitem{PLB2}
M. Znojil and H. B. Geyer, Phys. Lett. B 649 (2007) 494
(erratum).

\bibitem{PLB3}
M. Znojil, Phys. Lett. B 647 (2007) 225.

\bibitem{determ}
M. Znojil, Phys. Lett. A 367 (2007) 300.

\bibitem{maximal}
M. Znojil, J. Phys. A: Math. Theor. 40 (2007) 4863.

\bibitem{II}
M. Znojil, J. Phys. A: Math. Theor. 40 (2007), to appear
(arXiv:0709.1569).

\bibitem{condit}
M. Znojil, Phys. Lett. B 650 (2007) 440.

\bibitem{selfdual}
M. Znojil, Broken spectral reflection symmetry in a solvable
PT-symmetric model, submitted (arXiv:0710.0457).

\bibitem{PTSQM}
C. M. Bender, S. Boettcher and P. N. Meisinger, J. Math. Phys. 40
(1999) 2201.

\bibitem{Shifman}
G. Dunne and M. Shifman,
Annals of Physics 299 (2002) 143.

\bibitem{BG}
V. Buslaev and V. Grecchi, J. Phys. A: Math. Gen. 26 (1993) 5541.


\bibitem{Cannata}
C. M. Bender and K. A. Milton,
%
Phys. Rev. D 57 (1998) 3595;

A. A. Andrianov, F. Cannata, J-P. Dedonder and M. V. Ioffe,
Int. J. Mod. Phys. A 14 (1999) 2675;

M. Znojil, F. Cannata, B. Bagchi and R. Roychoudhury,
Phys. Lett. B 483 (2000) 284;

A. Mostafazadeh, Nucl. Phys. B 640 (2002) 419;

M. Znojil,
J. Phys. A: Math. Gen. 35 (2002) 2341;

G. L\'evai, Czech. J. Phys. 54 (2004) 1121;

B. F. Samsonov and V. V. Shamshutdinova, J. Phys. A: Math. Gen. 38
(2005) 4715;

B. Bagchi, A. Banerjee, E. Caliceti, F. Cannata, H. B. Geyer, C.
Quesne and M. Znojil, Int. J. Mod. Phys. A 20 (2005) 7107;

A. Gonzalez-Lopez and T. Tanaka, J. Phys. A: Math. Gen. 39 (2006)
3715;

T. Curthright, L. Mezincescu and D. Schuster, J. Math. Phys., to
appear (arXiv: quant-ph/0603170).

\bibitem{BM}  C. M. Bender and K. A. Milton,
Phys. Rev. D 55 (1997) R3255;

F. Kleefeld, Czech. J. Phys. 55 (2005) 1123;

V. Jakubsk\'y and J. Smejkal, Czech. J. Phys. 556 (2006) 985.

\bibitem{Alifirst}
  A. Mostafazadeh,
Class. Quantum Grav. 20 (2003) 155 and Czech. J. Phys. 54 (2004)
93;

A. A. Andrianov, F. Cannata and A. Y. Kamneschchik, J. Phys. A:
Math. Gen. 39 (2006) 9975.

\bibitem{BB}
C. M. Bender and S. Boettcher, Phys. Rev. Lett. { 80} (1998) 5243.

\bibitem{BBb}
F. M. Fern\'{a}ndez, R. Guardiola, J. Ros and M. Znojil, J. Phys.
A: Math. Gen. 31 (1998) 10105.

F. Cannata, G. Junker and J. Trost, Phys. Lett. {\ A 246} (1998)
219 (quant-ph/9805085);

E. Delabaere and F. Pham, Phys. Letters A 250 (1998) 25.

\bibitem{solvable}
M. Znojil, 
Phys. Lett. A 259 (1999) 220;

B. Bagchi and R. Roychoudhury, J. Phys. A: Math. Gen. 33 (2000)
L1;

G. Levai and M. Znojil,
 J. Phys. A: Math. Gen. 33
(2000) 7165.

\bibitem{BBJ}
C. M. Bender, D. C. Brody and H. F. Jones, Phys. Rev. Lett. 89
(2002) 0270401;

R. Kretschmer and L. Szymanowski, Czech. J. Phys. 54 (2004) 71;

A. Mostafazadeh, Czech. J. Phys. 54 (2004) 1125.

\bibitem{turek}
Q. Wang, Czech. J. Phys. 54 (2004) 143;

A. Mostafazadeh and S. Ozcelik,  Turk. J. Phys. 30 (2006) 437;

E. M. Gr{ae}fe and H. J. Korsch, Czech. J. Phys. 56 (2006) 1007.

\end{thebibliography}
\end{document}